\documentclass[cits]{PoS}

\title{Are the distances of galactic microquasars reliable?}

\ShortTitle{Are the distances of galactic microquasars reliable?}

\author{\speaker{C\'edric Foellmi}\\
        European Southern Observatory, 3107 Alonso de Cordova, Casilla 19, Vitacura, Santiago, Chile \\
        E-mail: \email{cfoellmi@eso.org}\\
        Laboratoire d'Astrophysique, Observatoire de Grenoble, 414 rue de la Piscine, 38400 Saint-Martin d'H\`eres, France\\
        E-mail: \email{cfoellmi@obs.ujf-grenoble.fr}}

\abstract{We review some specific and quantitative issues in popular methods used to determine the distance of galactic objects, in the context of the microquasars GRO J1655-40 and 1A 0620-00. In particular, we discuss the domain of validity of the relationship between the Sodium lines and the color excess, and the supposedly systematic overestimation of optical absorption from X-ray data. Both problems imply a possible underestimation of the absorption toward GRO J1655-40. 

We also discuss the main issue concerning the absolute magnitude of the secondary star in our own maximum-distance method, that has been used on GRO J1655-40. We show in this respect that the dynamical parameters of this microquasar are not definitive, and that the mean properties of the secondary star is not so much different from that of a normal single star of same spectral type. 

Following the possibility that GRO J1655-40 is located at about 1.0 kpc, associated with the open cluster NGC 6242, we rise the question: which black-hole is the closest to the Sun? In that context, we quickly discuss the situation of the microquasar 1A 0620-00, by addressing the specific point of its 30-years-old extinction value.

Finally, we present some details about a possible new distance method, using the surface properties of the stars, based on the cross-correlation function bisector, and that is now accessible thanks to the advent of a high-resolution echelle spectrograph called CRIRES at the VLT. We quickly discuss the strengths and weaknesses of this method, and how it possibly address the caveats of the other methods.

}

\FullConference{VI Microquasar Workshop: Microquasars and Beyond\\
		September 18-22 2006\\
		Societ\`a del Casino, Como, Italy}

\begin{document}

%------------------------------------------------------------------------------------
\section{Why is the distance issue relevant?}

The issue of the distance of microquasars in our galaxy, or even beyond, is obviously relevant in the sense that it is a very basic parameter of the models. In practice however, the distance values used in these models or simply spread in the literature,  are not always reliable, sometimes by a large factor. It may be worth to recall that even the distance of GRS 1915+105 is somewhat subject to caution, even if most of the studies (actually all but one: \cite{Kaiser-etal-2004}) converge to a single and consensual value of about 11~kpc.\footnote{Since this specific contribution is also about challenging a supposedly robust distance value "confirmed" by many studies, we prefer to reserve our judgement on that of GRS 1915+105. Moreover, a quick verification of the concordant studies about this microquasar show that they share most of their methods, requiring a dedicated study.} The answer to the title being known, it is nonetheless interesting to discuss the details to reveal where are the weaknesses and how to address them.

Although many of the remarks below can be understood in a very general context, we discuss the distance problems encountered with the galactic microquasars GRO J1655-40 and 1A 0620-00. The distance issue of the former is relevant for at least three reasons. First, GRO J1655-40 was at the time of its discovery, the second galactic microquasar (just after GRS 1915+105) showing superluminal radio jets \cite{Hjellming-Rupen-1995}. This property, and consequently the way the engine powering these jets is modeled, is directly related to the distance. Second, the distance of GRO J1655-40 is relevant in the context of finding the closest black-hole to the Sun. This "fashionable" question is however of first interest for other modes of observations, because if close enough, the elected black-hole -- which might be 1A 0620-00 -- could become an accessible target for the Very Large Telescope Interferometer and, in the late future, a privileged one for space missions dedicated to black-hole physics such as the European satellite XEUS. Finally, the distance is important to understand the context of formation of a microquasar (see \cite{Foellmi-etal-2006b}), and its possible trajectory in the galaxy, as shown already in \cite{Mirabel-etal-2002}.

%------------------------------------------------------------------------------------
\section{The distance of GRO J1655-40}

Part of the discussion below concerning GRO J1655-40 is already published in \cite{Foellmi-etal-2006b}, where we developed a new maximum-distance method. Using new VLT-UVES spectra, a comparison with nearby stars led us to find an upper limit of $D<1.7$ kpc, which contradicts the often quoted distance of $D=3.2\pm0.2$ kpc determined in \cite{Hjellming-Rupen-1995} from a kinematic model of the radio-jets. Following \cite{Mirabel-Rodriguez-1994}, we have already shown that this model can provide an {\it upper limit} only on the distance: $D_{\textrm{{\scriptsize max}}} < 3.53$ kpc. As for the lower limit of 3 kpc, it is given by the radio spectrum presented in \cite{Tingay-etal-1995} and is also subject to caution. The details concerning this specific distance value have been presented in \cite{Foellmi-2006c}.

%------------------------------------------------------------------------------------
\subsection{The unsolved problem of optical absorption}

The issue of the absorption is originating from the need to correctly estimate the difference between the apparent and absolute magnitudes, which leads directly to the distance through the well-known relation:
\begin{equation}
m_{\textrm{{\scriptsize obs}}} - M - A = 5 \log(D) - 5
\label{magdist}
\end{equation}
where $m_{\textrm{{\scriptsize obs}}}$ is the observed apparent magnitude ($m_{\textrm{{\scriptsize obs}}}$ = $m_{\textrm{{\scriptsize true}}} + A$), $M$ the absolute magnitude, $A$ the absorption and $D$ the distance in parsecs. In microquasars, the absolute magnitude of the secondary is usually obtained through the combination of the spectral type (which provides an estimate of the effective temperature $T_{\textrm{{\scriptsize eff}}}$) and the orbital dynamics (which provides the semi-major axis). Assuming that the secondary star is filling its Roche lobe, and using the Stefan-Boltzman law, it is possible to compute the intrinsic luminosity and hence the absolute magnitude of the secondary star. This specific point for GRO J1655-40 is discussed below. 

Things are getting problematic when one tries to estimate the absorption, which is directly related to the color excess $E(B-V)$ by the standard formula $A_V = R \times E(B-V)$ and where $R \approx 3.1$. The relationship between equivalent width of Sodium lines and the color excess is often used, particularly in the case of the distance of GRO J1655-40 (e.g. \cite{Bianchini-etal-1997,Hynes-etal-1998}). One could add however that the validity range of this relationship is very small. Following \cite{Munari-Zwitter-1997}, the equivalent widths must be comprised between 0.13 and 0.75\AA, which implies in turn a color excess $E(B-V)$ between 0. and 0.4. Using gaussian profiles for the absorption lines, and a minimal 2.5 pixel resolution element, the validity range translates to a minimal spectroscopic resolving power of 8000 for the broadest lines, and 50 000 for the narrowest (this latter value being reached for instance by the echelle spectrograph UVES at the VLT). 

The estimated color excess of GRO~J1655-40 (see e.g. \cite{Hynes-etal-1998}) of 1.3 is by far outside this range. Moreover, as clearly shown in \cite{Foellmi-etal-2006b}, high-resolution spectra revealed the saturated and multi-profiled nature of these lines in GRO J1655-40, making them totally unusable. It also means that the absorption towards this microquasar is possibly underestimated.

%------------------------------------------------------------------------------------
\subsection{The (supposedly) systematic overestimation of absorption from X-ray data.}

Another mean of estimating the optical absorption consists of measuring the column density of hydrogen with X-ray data (under the -- generally reasonable -- assumption of cospatiality of the hydrogen and interstellar dust). However, in their detailed study of the distances of X-ray binaries, \cite{Jonker-Nelemans-2004} claimed that the absorption measured from X-rays is systematically overestimated compared to that obtained from optical data (which often relies on the equivalent widths of Sodium lines mentioned above). They presented a table of 14 sources for the comparison (see their Table~3).

However among these 14 sources, four targets have a missing value in one of the two bands (GS~1009-45, XTE~J1118+480, H~1705-250 and SAX~J1819.3-2525), five have consistent values between the two bands given the quoted uncertainties (GRO~J0422+32, 1A~0620--00, GRO~J1655--40, GX~339--4 and GS~2~Section 023+338), leaving 5 systems only for a true comparison: GS~1124--684, 4U~1543--47, XTE~J1550--564, XTE~J1859+226 and GS~2000+25. For the X-ray binary XTE~J1550--564, \cite{Sanchez-Fernandez-etal-1999} obtained a color excess $E(B-V)=0.7$ value from equivalent width of the Sodium lines of 2.4\AA\, which is much larger than the upper limit of the validity range defined above. Moreover, they use a low-resolution (2\AA) spectrum, and it is not clear if the lines are saturated or not, as it was the case for GRO J1655-40 \cite{Foellmi-etal-2006b}. As for GS~2000+25, the value from X-ray observations quoted by \cite{Jonker-Nelemans-2004} is $A_V = 6.4\pm1.0$, citing \cite{Tsunemi-etal-1989}. However, in this latter paper, the value quoted is $A_V = 4.4$ (or $\log N_H = 22.06\pm0.006$), said to be in agreement with the optical estimation by \cite{Chevalier-Ilovaisky-1990} that is the reference given by \cite{Jonker-Nelemans-2004} for the optical value!

In summary, the {\it systematic} overestimation of the absorption from X-ray observations is not established.  One might even wonder about a possible systematic {\it underestimation} of the absorption from {\it optical} data. In this context, we cannot exclude the result of \cite{Greiner-etal-1995} on GRO J1655-40 using $ROSAT$ data and who find an absorption of $A_V = 5.6$ mag.

%------------------------------------------------------------------------------------
\section{The maximum-distance method of Foellmi et al. (2006)}

%- - - - - - - - - - - - - - - - - - - - - - - - - - - - - - - - - - - - - - - - - - 
\subsection{Quick presentation}

We have developed and presented a new maximum-distance method (\cite{Foellmi-etal-2006b}), applied it to GRO J1655-40 and found an upper limit to the distance of 1.7 kpc. This method is based on the comparison of the spectroscopic flux of the microquasar {\it in quiescence} with that of a nearby single star of same spectral type. From archival VLT-UVES spectra, we determined a F6IV spectral type for the secondary star, in agreement (within one spectral type) with previous works. After a verification with two different methods that our spectra were not showing any detectable contribution from the accretion disk, the mean flux-calibrated spectrum of this microquasar was then compared to that of the F6IV star HD~156098, using a spectrum taken with the exact same configuration, and available in the UVES POP\footnote{{\tt http://www.sc.eso.org/santiago/uvespop}} (see \cite{Bagnulo-etal-2003}). Then, we used the following relation:
\begin{equation}
a = 5 \, \log \left( \frac{D_2}{D_1} \frac{1}{\sqrt{f}} \right) + M_2 - M_1 \geq 0
\end{equation}
where $a$ designate the absorption\footnote{See \cite{Foellmi-etal-2006b} for a discussion on why $a$ does not mean exactly the normal broad-band photometric absorption.}, $D_2$ and $D_1$ the distance of the nearby star and GRO J1655-40 respectively, $f$ the spectroscopic flux ratio, and $M$ the absolute magnitudes similarly to the distances. Let first emphasize that we claimed that the value of 1.7 kpc obtained through the comparison with five nearby stars of similar spectral types is a strong upper limit because it is based on the extreme case where $a = 0$. Obviously, the absorption {\it is} certainly larger than zero. Let emphasize as well that it is not a distance method, but is providing instead an {\it upper limit} only to the distance. In that sense, it is not a {\it measurement} of the distance.
 
This method is based on a strong assumption about the absolute magnitude of the secondary star of the microquasar. We chose to bracket the possible value of $M_1$ to that of $M_2$ more or less one magnitude, i.e. $M_1 = M_2 \pm 1$, expecting naturally that the true value of $M_1$ is comprised in this range.  

%- - - - - - - - - - - - - - - - - - - - - - - - - - - - - - - - - - - - - - - - - - 
\subsection{The issue of the absolute magnitude}

This distance method is problematic because the absolute magnitude $M_1$ is systematically uncertain. In other words, the assumption might or might not be verified,  and there is no obvious way to know if it is the case or not. On the other hand, it is generally assumed that in low-mass X-ray binaries, there is no stellar wind strong enough to create and sustain an accretion disk, and that only Roche-lobe overflow from the secondary star is able to do so. Consequently, the secondary star, even in quiescence, should have an effective radius equal to, or at least similar to, the radius of its Roche lobe. 

Although all studies of GRO J1655-40 agree on the spectral subtype, the value of the luminosity obtained by \cite{Orosz-Bailyn-1997} or that of \cite{Hynes-etal-1998} ($M_V=0.7$) is very different from the absolute magnitude of a normal F6IV star: $M_V=3.2$, as given by \cite{Gray-1992} (which is the reference given by \cite{Hynes-etal-1998}). This is considered as a clear evidence in favor of the presence of special conditions implying that the comparison between the secondary star and a single star of same spectral type and luminosity class is flawed. Consequently, the assumption of \cite{Foellmi-etal-2006b} should also be flawed.

But a change from 3.2 to 0.7 implies a change of radius by a factor of three! And the change of radius implies a significant change of surface gravity, which contradicts the agreement between the UVES spectrum of GRO J1655-40 in quiescence, and that of the  HD 156098, as shown by \cite{Foellmi-etal-2006b}: $T_{\textrm{{\scriptsize eff}}}$ = 6480$K$, $\log g$=3.94, Fe/H=0.09 (see also \cite{Edvardsson-etal-1993}). More interestingly, HD 156098 has a known \emph{Hipparcos} distance and its absolute magnitude is thus available: $M_V=2.0\pm0.2$. This brighter magnitude can be considered in two ways. Either expected "special conditions" justifying the absence of comparison of the stars still holds for some reasons, and the effect on the radius being not {\it so} large (but still a factor two), it becomes explainable by some effects, such as irradiation (see e.g. \cite{Phillips-Podsiadlowski-2002}). Or the star in GRO J1655-40 is in fact not so much different (see also below) from a normal F6IV star. In both cases, the validity of the assumption of \cite{Foellmi-etal-2006b} is becoming more reasonable.

Of course, remains the problem of the Roche radius and the fact that the star should fill it. But are the dynamical parameters of GRO J1655-40 well established?

%- - - - - - - - - - - - - - - - - - - - - - - - - - - - - - - - - - - - - - - - - - 
\subsection{Is the orbital dynamic of GRO J1655-40 well constrained?}

Let us compare two important studies about the dynamics of GRO J1655-40: \cite{Orosz-Bailyn-1997} and \cite{Beer-Podsiadlowski-2002}\footnote{\cite{Beer-Podsiadlowski-2002} refer mostly to the results of \cite{Greene-etal-2001} for a comparison of their models. The results of the latter are in fact in good agreement with that of \cite{Orosz-Bailyn-1997} and are thus not reproduced.}. The parameters are summarized in Table~\ref{orbital-parameters}. In addition we have computed the true equatorial rotational velocity of the secondary star from the inclination angle $i$, the Roche effective radius using the formula of \cite{Paczynski-1971} (see also \cite{Jonker-Nelemans-2004}) from the period $P$ and the secondary mass $M_2$, and then the equatorial velocity of the secondary star if it had the same radius. Finally, from the luminosities we computed the corresponding absolute magnitude. For the projected rotational velocity, we used the value of \cite{Greene-etal-2001} ($V_{\textrm{{\scriptsize rot}}} \sin i$=93 km\,s$^{-1}$) suffering no effect from tidal distortion since it has been measured at phase 0, and with which \cite{Foellmi-etal-2006b} agree perfectly.

%\begin{tabular}{lr@{\,$\pm$\,}lr@{\,$\pm$\,}lr@{\,$\pm$\,}l} \hline
%Parameter & \multicolumn{2}{c}{\cite{Orosz-Bailyn-1997}} & \multicolumn{2}{c}{\cite{Greene-etal-2001}} & \multicolumn{2}{c}{\cite{Beer-Podsiadlowski-2002}} \\ \hline
%Orbital Period [$days$]& 2.62157 & 0.00015 & 2.62191 & 0.0002 & \multicolumn{2}{c}{X} \\
%Mass Ratio $q$ & 3.0 & 0.4 & 2.6 & 0.3 & 3.9 & 0.6 \\ 
%Inclination $i$ [$^{\circ}$]  & 69.8 & 0.8 & 72.2 & 1.9 & 68.65 & 1.6 \\
%$T_{\textrm{{\scriptsize pole}}}$ [$K$]& \multicolumn{2}{c}{6768} & 6575 & 375 \\
%Mass $M_1$ [$M_{\odot}$] & 7.0 & 0.6 & 6.3 & 0.5 & 5.40 & 0.30 \\
%Mass $M_2$ [$M_{\odot}$] & 2.4 & 0.5 & 2.4 & 0.4 & 1.45 & 0.35 \\
%Luminosity $L_2$ [$L_{\odot}$]& 46.6 & 13.6 & \multicolumn{2}{c}{31.9 -- 40.6} & 21.0 & 6.0 \\
%$T_{\textrm{{\scriptsize eff}}}$ [$K$] & \multicolumn{2}{c}{6500 (fixed)} & \multicolumn{2}{c}{6336 (fixed)}& 6150 & 350 \\
%Color Excess $E(B-V)$ & 1.3 & 0.1$^{\mathbf{a}}$ & \multicolumn{2}{c}{1.076} & 1.0 & 0.1 \\ \hline
%$V_{\textrm{{\scriptsize rot}}}$ ($V_{sin\textrm{{\scriptsize i}}}$=93) [km\,s$^{-1}$] &  99.1 & 0.1 & 97.7 & 1.0 & 99.8 & 1.0 \\
%Roche Radius [$R_{\odot}$] & 4.95 & 0.15 & 4.9 & 0.3 & 4.2 & 0.2 \\
%Equatorial Roche Velocity [km\,s$^{-1}$] & 95.7 & 6.7 & 95.7 & 5.3 & 80.9 & 6.6 \\ 
%Absolute Magnitude $M_V$ [mag] & 0.7 & 0.3 & \multicolumn{2}{c}{0.81 -- 1.08} & 1.5 & 0.3 \\ 
%Quoted Distance [kpc] & 3.2 & 0.2$^{\mathbf{b}}$ & \multicolumn{2}{c}{3.8$^{\mathbf{c}}$} & 3.2 & 0.2$^{\mathbf{d}}$ \\ 
%Computed Distance [kpc] & 3.0 & 1.0 & 3.9 & 0.4 & 3.3 & 0.9$^{\mathbf{b}}$ \\ \hline
%\end{tabular}

\begin{table}
\centering
\begin{tabular}{lr@{\,$\pm$\,}lr@{\,$\pm$\,}l} \hline
Parameter & \multicolumn{2}{c}{\cite{Orosz-Bailyn-1997}} & \multicolumn{2}{c}{\cite{Beer-Podsiadlowski-2002}} \\ \hline
Orbital Period $P$ [$days$]& 2.62157 & 0.00015 & \multicolumn{2}{c}{X} \\
Mass Ratio $q$ & 3.0 & 0.4 & 3.9 & 0.6 \\ 
Inclination $i$ [$^{\circ}$]  & 69.8 & 0.8 & 68.65 & 1.6 \\
Mass $M_1$ [$M_{\odot}$] & 7.0 & 0.6 & 5.40 & 0.30 \\
Mass $M_2$ [$M_{\odot}$] & 2.4 & 0.5 & 1.45 & 0.35 \\
Luminosity $L_2$ [$L_{\odot}$]& 46.6 & 13.6 & 21.0 & 6.0 \\
Temperature $T_{\textrm{{\scriptsize eff}}}$ [$K$] & \multicolumn{2}{c}{6500 (fixed)}& 6150 & 350 \\
Color Excess $E(B-V)$ & 1.3 & 0.1$^{\mathbf{a}}$ & 1.0 & 0.1 \\ \hline
$V_{\textrm{{\scriptsize rot}}}$ ($V_{sin\textrm{{\scriptsize i}}}$=93) [km\,s$^{-1}$] &  99.1 & 0.1 & 99.8 & 1.0 \\
Roche Radius [$R_{\odot}$] & 4.95 & 0.15 & 4.2 & 0.2 \\
Equatorial Roche Velocity [km\,s$^{-1}$] & 95.7 & 6.7 & 80.9 & 6.6 \\ 
Absolute Magnitude $M_V$ [mag] & 0.7 & 0.3 & 1.5 & 0.3 \\ 
%Quoted Distance [kpc] & 3.2 & 0.2 & 3.2 & 0.2 \\ 
%Computed Distance [kpc] & 3.0 & 1.0 & 3.3 & 0.9 \\ \hline
\end{tabular}
\caption{Comparison table between various orbital parameters of GRO J1655-40 as measured by \cite{Orosz-Bailyn-1997} and modeled by \cite{Beer-Podsiadlowski-2002}. It illustrates that the dynamics of GRO J1655-40 is not yet fully understood. The values of mass ratio, inclination angle and masses of \cite{Orosz-Bailyn-1997} are taken from their Table~7, using the $3\sigma$ values. Notes: $^{\mathbf{a}}$ The value of the color excess (that is used to compute also the observed luminosity), is taken from \cite{Horne-etal-1996} (but see text).}
\label{orbital-parameters}
\end{table}

Clearly, the two studies give different results about the secondary star in GRO J1655-40. In particular, we note in \cite{Beer-Podsiadlowski-2002} the large decrease of luminosity by a factor $\sim$2, and a smaller mass of the secondary star by 40\%! Their absolute magnitude is close to agreeing with that of the star HD 156098 mentioned above. Interestingly they also obtain a much smaller Roche radius, which makes the rotational velocity significantly larger than the computed equatorial velocity at the surface of a sphere whose radius is equal to the effective Roche radius. Does it means that the star is over-synchronous, while \cite{Orosz-Bailyn-1997} (see also \cite{Greene-etal-2001}) find a synchronization? 

These results unfortunately cannot be considered as reliable for obtaining all the necessary ingredients of Eq.~\ref{magdist} to compute the distance. As a matter of fact, the luminosity of \cite{Orosz-Bailyn-1997} is obtained with two quantities that are unreliable: the distance of 3.2~kpc from \cite{Hjellming-Rupen-1995} and the color excess from an IAU Circular: \cite{Horne-etal-1996}. This Circular uses $HST$ data that are obviously not presented but could not be found elsewhere neither. As for \cite{Beer-Podsiadlowski-2002}, their results are also based on the distance of 3.2~kpc of \cite{Hjellming-Rupen-1995}. They mention that they do find a solution with their model with a much smaller distance, but they discarded it, taking 3.2~kpc as a firm constraint. They also argue that their results must be consistent, since they found a color excess in agreement with all other studies. We have shown above (and in \cite{Foellmi-etal-2006b}) how the extinction is certainly underestimated.

We note that it is tempting to use value of the absolute magnitude of \cite{Beer-Podsiadlowski-2002}, with the apparent magnitude $m_V=17.12$ of \cite{Orosz-Bailyn-1997} and the absorption from X-ray data by \cite{Greiner-etal-1995} ($A_V = 5.6$) which leads to a distance of 1.01~kpc, consistent with the possibility that GRO J1655-40 is originating from the open cluster NGC 6242 (see \cite{Foellmi-etal-2006b,Mirabel-etal-2002}). But only a consistent recomputation of the model of \cite{Beer-Podsiadlowski-2002} would be acceptable. 

In summary, taking some results of \cite{Beer-Podsiadlowski-2002}, it seems that the {\it mean} properties of the secondary star of GRO J1655-40 are not so much different from that of the single and nearby F6IV star HD 156098. There respective absolute magnitudes are actually consistent within one magnitude, implying that the assumption on the absolute magnitudes in our maximum-distance method is valid, at least in the case of GRO J1655-40.

%------------------------------------------------------------------------------------
\section{The distance of 1A 0620-00}

It has shown above that the distance of GRO J1655-40 found in the literature is certainly unreliable, and that the combination of an underestimation of the absorption with a probable fainter absolute magnitude leads to the possibility that GRO J1655-40 might be much closer \cite{Foellmi-etal-2006b}. If indeed GRO J1655-40 is originating from the open galactic cluster NGC 6242 located at 1.0 kpc, as its proper motion vector seem to indicate (see \cite{Mirabel-etal-2002}), it competes with the microquasar 1A 0620-00 for being the closest black-hole to the Sun. 

The distance of 1A 0620-00 is however very uncertain as well. For instance \cite{Jonker-Nelemans-2004}, \cite{Gelino-etal-2001a} and more recently \cite{Gallo-etal-2006} use a 30-years-old value of the extinction used to estimate the absorption. This extinction value is quoted from \cite{Wu-etal-1983} in 1983 who do nothing but quote their own results \cite{Wu-etal-1976} in 1976. This latter paper describes how the authors determine the extinction by nulling the feature seen at 2200\AA\ in their UV spectrum. A simple verification of \cite{Wu-etal-1976} reveals however that the UV spectrum is actually made of five points only, since the Astronomical Netherlands Satellite used to obtain this spectrum had only 5 channels. Most interestingly, a HST/STIS spectrum has been published of this source \cite{McClintock-Remillard-2000} which shows the total absence of this feature at 2200\AA. We must note however, that given the direction of 1A 0620-00 in the Galaxy ($l=209.96^{\circ}$, $b=-6.54^{\circ}$), usual extinction methods will certainly not be valid at all. 

The total absence of the feature at 2200\AA\ is pointing toward the possibility that 1A 0620-00 is significantly closer to the Sun than 1.0 kpc. A quick application of our maximum-distance method \cite{Foellmi-etal-2006b} on this object, using the UVES spectrum of HD 209100, also points toward a smaller distance from the Sun \cite{Foellmi-etal-2006e}: $D\sim 200$ pc. {\it However} we have not done yet a modeling of the spectrum as for GRO J1655-40, and this value must be considered as an indication only.

%------------------------------------------------------------------------------------
\section{A new distance method base on surface properties of stars?}

As shown in \cite{Foellmi-etal-2006b}, there are many distance methods that have been used in the case of GRO J1655-40 and 1A 0620-00. Of course, none of these are perfect. Waiting for the future european satellite \emph{GAIA} to solve the issue, one could try to address the caveats of previous method. While it remains difficult to find a solution to the problem of the extinction, the combination of the temperature from the spectral type and an uncertain Roche radius is relying on many observational difficulties. Moreover, the temperature of the secondary star changes not only with the orbital period, but also across the stellar surface \cite{Beer-Podsiadlowski-2002} and the possible irradiation is usually not taken into account when one measures the radial-velocities (see \cite{Phillips-Podsiadlowski-2002}). We briefly describe here a totally new distance method, already quickly presented elsewhere \cite{Foellmi-2006d}.

This new distance method is based on the combination of two recent results. First, \cite{Gray-2005} has shown a relationship between the height of the bluemost point of the single-line bisector\footnote{The bisector of a spectroscopic line is the profile of the curve composed of the central midpoints between the two wings of the lines. See \cite{Gray-2005} for a picture. It has a classical "C" shape.} of late-type stars, of very different luminosity classes. As  such, it could already be used as distance method, since it allows to obtain the absolute magnitude independently. This relationship is actually a consequence of how stellar atmospheric lines form, and depends on the granularity of the surface (see e.g. \cite{Asplund-etal-2000}). However, to obtain a usable single-line bisector on a spectrum, one requires a very high resolving power (at least 50 000), and a very high Signal-to-Noise ratio (at least S/N $\geq 300-500$). This  prevents this possible method to be used on fairly faint stars, since only the brightest stars in the sky are accessible.

On the other hand, \cite{Dall-etal-2006} have studied the bisectors of the cross-correlation function (CCF) of spectra taken with the one-meter-per-second accuracy spectrograph HARPS installed at the ESO 3.6m telescope in La Silla Observatory (Chile), and which is extensively used to search for extrasolar planets. \cite{Dall-etal-2006} have shown that the bisector of the CCF can be used as much the same way as the single-line bisector. Even better, they show a relationship (for a small sample of stars though) between the bisector parameters and the absolute magnitude, and this even for the hotter stars for which the bluemost point of the bisector disappear.

CCF bisectors are much more accessible than single-line bisector, since they can be seen as a "mean line profile", and do not require the extremely high S/N if enough lines can be used. However, for objects like GRO J1655-40, it is still difficult to obtain such bisector in the optical (V$\sim$18), and a quick calculation show that two weeks of continuous 24-hours observing time of GRO J1655-40 with UVES are necessary to provide the required spectrum quality. This is why it is necessary to observe in the near infrared (NIR), where microquasars are much brighter (GRO J1655-40: K$\sim$12--13). As a very timely opportunity, a high-resolution cryogenic NIR echelle spectrograph called CRIRES is being commissioned nowadays at the Very Large Telescope. A program is already ongoing on this instrument to check the validity of this relationship, and its application to microquasars.

There are various identified problems with this method. First, the relation as found by \cite{Dall-etal-2006} is not yet fully understood. Moreover, it has been calibrated over a small sample of stars only, and in the optical domain. The classical "C" shape of single-line bisector is also expected to be of smaller amplitude at longer wavelengths, and it is not clear how this will appear in the NIR with CRIRES. On the other hand, this method certainly deserves a detailed exploration since it potentially gives access to the absolute magnitude (even a 10-20\% accuracy is already a significant improvement for microquasars) without the need to combine the dynamics of the system to have the radius and spectral type to have the temperature of the secondary star. 

One might argue that the apparent properties of the surface of an (irradiated) star are changing with the orbital phase, and are probably different from a normal single and standalone star. However, both problems can be avoided by observing at phase 0.  If the period is short, the superposition of multiple spectra at the same phases might be required. We note nonetheless that the relationship found by \cite{Dall-etal-2006} seems still valid for stars slightly above the granularity boundary. 

%------------------------------------------------------------------------------------
\section{Conclusion}

In this contribution, we have emphasized again that the extinction problem is a complicated one, and probably unsolved. We have shown that there is a possible systematic underestimation of the absorption from optical data towards GRO J1655-40, and  that the supposedly systematic overestimation of the absorption from X-ray data is not established. It means that the results of \cite{Greiner-etal-1995}, who found from $ROSAT$ data an absorption toward this microquasar much larger compared to any other optical value, cannot be discarded.

We have tried to address the problem of the assumption in our maximum-distance method \cite{Foellmi-etal-2006b}. In that respect, we have shown that the expected special conditions preventing to compare the secondary star in GRO J1655-40 with a normal single star of same spectral type and luminosity class, must actually provide an explanation for a very large change of radius (at least a factor two). On the other hand, we also questioned the relevance of these effects by showing that the dynamics of the system is by far not fully understood. In summary, the validity of the assumption of our maximum-distance method is emphasized, since the properties of GRO J1655-40 are getting closer to that of its comparison star HD 156098.

Finally, we have shown that the extinction toward 1A 0620-00 relies on an 30-years-old measurement in a UV "spectrum", that appears to be totally unreliable. The distance of this microquasar is under investigation by our team using the promising new distance method presented here, using CRIRES bisectors. This method could  possibly address the caveat of relying to the spectrum and the dynamics to estimate the absolute magnitude of the secondary star of a microquasar.

%------------------------------------------------------------------------------------
\acknowledgments

The work on the distance methods could not have been made without the great help of two spectroscopist experts. I warmly acknowledge again the friendly collaboration with T.H. Dall and E. Depagne. I also thank D. Dravins for pointing out some difficulties inherent to the bisector method. 

\bibliography{/Users/cedric/science/literature/mainbiblio}
\bibliographystyle{PoS}

\end{document}